\documentclass[twocolumn,english,prl,showpacs,superscriptaddress]{revtex4}
\usepackage{times}
\usepackage[T1]{fontenc}
\usepackage[latin1]{inputenc}
\usepackage{graphicx}
\usepackage{amssymb}

\makeatletter

\usepackage{babel}
\makeatother
\begin{document}

\title{Single Molecule Chemical Reaction: Kramers Approach Revisited}

\author{G. Margolin}

\affiliation{Department of Chemistry and Biochemistry, Notre Dame University,
Notre Dame, IN 46556}

\author{E. Barkai}

\affiliation{Department of Chemistry and Biochemistry, Notre Dame University,
Notre Dame, IN 46556}

\affiliation{Department of Physics, Bar Ilan University, Ramat Gan, Israel 52900}

\date{\today{}}

\begin{abstract}
Single molecule chemical reactions yield new insight into fluctuation
phenomena which are obscured in measurement of ensemble of molecules.
Kramers escape problem is investigated here in a framework suitable
for single molecule reactions. In particular we obtain distributions
of escape times in simple limiting cases, rather than their mean,
and investigate their sensitivity on initial conditions. Rich physical
behaviors are observed: sub-Poissonian statistics when the dynamics
is only slightly deviating from Newtonian, super-Poissonian behavior
when diffusion is dominating, and Poissonian behavior when Kramers
original conditions hold. By varying initial conditions escape time
distributions can follow a (usual) exponential or a $\tau^{-3/2}$
decay, due to regular diffusion. We briefly address experimental results
which yield the $\tau^{-3/2}$ behavior (with cutoffs) and propose
that this behavior is universal.
\end{abstract}


\maketitle
Chemical reaction of a single molecule evolving between two states
$A\rightleftharpoons B$ or of two species $A+B\rightleftharpoons AB$
is now followed in many laboratories using single molecule spectroscopy
techniques \cite{Xie02,BJS04,Lu98,Deniz00,Rhoades03,Rhoades04,Flomenbom05}.
Such experiments yield detailed statistical information on chemical
conformational changes, and simple chemical processes in condensed
phase environments, for example the distribution of occupation times
of states \emph{A} and \emph{B} in the process $A\rightleftharpoons B$.
Such information is impossible to obtain when measurements are made
on many molecules, since the ensemble averaging wipes out the detailed
information found on the single molecule level. For ensemble of molecules,
usually simple reactions are assumed to follow a rate process.

For ensemble of molecules undergoing a chemical reaction, classical
concepts like reaction coordinate, and rate equations, work well in
many cases. In particular Kramers model \cite{Kramers40,Hanggi90}
for activation over a barrier, is a fundamental tool for modeling
chemical reactions in condensed phase environments. In this manuscript
we analyze Kramers problem, in a framework of single molecules. We
first discuss Kramers original approach, and its limitations in the
single molecule domain. 

Kramers describes a chemical reaction in terms of a reaction coordinate
$x(t)$. The complicated interaction of the chemical species with
their environment is replaced with a stochastic one dimensional approach.
The coordinate $x(t)$ evolves in a deterministic force field $V(x)$,
and is also coupled to a thermal heat bath. The reaction coordinate
is supposed to escape a metastable state. The inverse of the average
time of escape $\left\langle \tau\right\rangle $, from the bottom
of the well, serves as an estimate on the ensemble averaged reaction
rate. Two important regimes of Kramers problem are the underdamped
and overdamped limits. Many refinements, non-trivial results and generalizations
of Kramers problem are known, e.g. Kramers turn over behavior, quantum
effects, and non-Markovian generalizations (e.g., \cite{Buttiker83,Melnikov86,Drozdov99};
see \cite{Hanggi90} for a review). Experimental validation of the
theory is also obtained \cite{McCann99}.

At least three aspects of Kramers problem must be revised in the context
of single molecule chemical reaction. The most obvious one is that
now we must consider the distribution of occupation times in a chemical
state, and not limit the theory to averaged escape rates. Previous
work considered temporal dependence of the rate \cite{Chen04,Bao04}
until it reaches an equilibrium value in a single escape (transition)
event, or the transient behavior after the particle injection close
to the bottom of the well \cite{Shneidman97}. The idea of fluctuating
rates in multiple transitions, in the context of single molecule experiments
has been scrutinized and used in, e.g., \cite{WangWolynes95,Chernyak99,Xie02,Brown03}.
A second issue is the sensitivity of single molecule chemical reactions
to initial conditions. Consider the ongoing chemical reaction $A\rightleftharpoons B$.
For example using fluorescence resonance energy transfer (FRET) methods,
one may follow the closing (state \emph{A}) and opening (state \emph{B})
of a large molecule \cite{Weiss99,Selvin00,Deniz00,Rhoades03,Rhoades04}.
The experimental data then yields the string of occupation times $\{\tau_{A}^{1},\tau_{B}^{2},\tau_{A}^{3},....\}$.
Following Kramers assume that such events are described by a reaction
coordinate, which goes over a single barrier, to cross from state
$A\rightarrow B$ and vice versa. Also we assume that a measurement
may distinguish between state $A$ and state $B$, as is shown in
many experiments. In terms of the reaction coordinate this means that
when $x(t)<x_{b}$ the system is in the state $A$, otherwise it is
in the state $B$ and $x_{b}$ is the boundary (usually and conveniently
assumed to be at the top of the potential barrier). Then immediately
after the transition event from say $A\rightarrow B$, the reaction
coordinate is in close vicinity to $x_{b}$. Thus a short time after
the transition from state $A\rightarrow B$ there is an increased
likelihood for a back transition $B\rightarrow A$. This possibility
is expected to yield bunching of chemical activity on the time axis,
i.e. to intermittency where strong activity is observed for some period
of time followed by periods of lesser activity. Of course to observe
such effects the resolution time in the experiments must be short
compared to relaxation times of the dynamics. A third important point
is that in single molecule experiments, an additional length scale
$\lambda$ is introduced into the problem. For example the radius
of the laser illumination field, or the F\"orster length scale in
FRET experiments \cite{Selvin00,Deniz00}. Roughly, in the reaction
$A\rightleftharpoons B$, a fluorescence signal is recorded if an
acceptor and donor are within a range of F\"orster radius from each
other (state \emph{A}) while otherwise the signal is zero (state \emph{B}).
In terms of the reaction coordinate this implies that when $x(t)<\lambda$
the system is in the state $A$ otherwise it is in the state $B$.
The point to notice is that in principle $\lambda$ can be anywhere
along the reaction coordinate. In particular the classical approach
of an escape from a bottom of the well over a maximum in the potential
field is not expected to be general. 

Hence we investigate Kramers escape problem, obtaining distribution
of escape times. With initial conditions both in vicinity of the escape
point and far from it. We classify the deviations from exponential
behaviors, and show that in many cases rate concept is not valid.
In particular we classify a turnover behavior from power law behavior
to exponential, and show that the power law behavior is general. 

\textbf{Model 1} We consider a classical particle undergoing underdamped
diffusion in a harmonic potential well. As well known the relevant
coordinate for underdamped motion is the action $I$. The probability
density function (PDF) \emph{P} of finding the particle with action
$I$ obeys \begin{equation}
\frac{\partial P}{\partial t}=\gamma\frac{\partial}{\partial I}\left(IP+\frac{2\pi k_{B}T}{\omega}I\frac{\partial P}{\partial I}\right)\label{Eq1}\end{equation}
 where $\omega$ is the harmonic frequency, $E$ is the particle energy,
$k_{B}T$ is the thermal energy, and $\gamma$ is the damping coefficient;
$\gamma\ll\omega$ (see Fig.%
\begin{figure}
\includegraphics[%
  width=0.70\columnwidth,
  keepaspectratio]{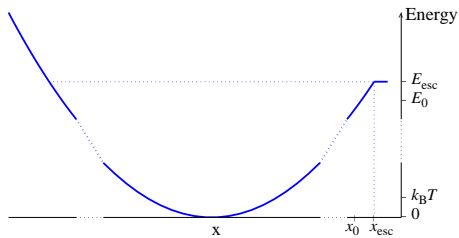}

\caption{\label{cap:Escape-def}Escape from a metastable well. In the underdamped
case, the relevant coordinate is the action $I$, which is related
to the energy $E$ of the particle by $I=2\pi E/\omega$. In the overdamped
case, the relevant coordinate is position x in real space.}
\end{figure}
\ref{cap:Escape-def}; $I=2\pi E/\omega$ for constant $\omega$).
In what follows we use dimensionless time $\tau=\gamma t$ and action
$\mathcal{I}=I\omega/2\pi k_{B}T$. Let $\phi(\tau)$ be the PDF of
escape times from $\mathcal{I}_{0}$ to $\mathcal{I}_{esc}>\mathcal{I}_{0}$.
Mathematically we use absorbing boundary conditions at $\mathcal{I}_{esc}$,
so that $\phi(\tau)$ is the first passage time distribution (FPTD)
from $\mathcal{I}_{0}$ to $\mathcal{I}_{esc}$. We obtained the Laplace
$\tau\rightarrow u$ transform of $\phi(\tau)$, using known solution
to Kummer's equation \cite{A&S,Siegert51}. We find \begin{equation}
\hat{\phi}(u)=\frac{_{1}F_{1}(u;1;\mathcal{I}_{0})}{_{1}F_{1}(u;1;\mathcal{I}_{esc})}\label{eq:Lphi-under}\end{equation}
 where $_{1}F_{1}(a;b;c)$ is the regular confluent hypergeometric
function. Since $_{1}F_{1}(0;b;c)=1$ the function $\phi(\tau)$ is
normalized to 1. We note that a second presentation of the solution
in terms of an infinite sum of exponentially decaying modes is possible,
in time domain. However, as we shall show now for an important parameter
regime, one cannot replace such standard eigenfunction solutions with
a summation over a finite number of modes consisting of a few exponential
functions, rather the solution exhibits a power law behavior. 

To quantify deviations from exponential statistics, we use the parameter\[
Q=\frac{\sigma^{2}}{\left\langle \tau\right\rangle ^{2}}-1,\]
where $\langle\tau\rangle$ is the average escape time and $\sigma^{2}=\left\langle \tau^{2}\right\rangle -\left\langle \tau\right\rangle ^{2}$
is its variance. If we have Poissonian behavior consistent with a
rate equation approach, then $Q=0$. When the dynamics is Newtonian
(i.e. diffusion is weak) the PDF of escape time is narrow, leading
to $Q<0$, a sub-Poissonian behavior. On the contrary, if the PDF
of escape times is wide spread, there is a super-Poissonian behavior
and $Q>0$. Using the small $u$ expansion of the exact solution,
we find\begin{equation}
\left\langle \tau\right\rangle =A(\mathcal{I}_{esc})-A(\mathcal{I}_{0})\label{eq:psi-mean}\end{equation}
and\begin{equation}
\sigma^{2}=A^{2}(\mathcal{I}_{esc})-A^{2}(\mathcal{I}_{0})-2[B(\mathcal{I}_{esc})-B(\mathcal{I}_{0})]\label{eq:psi-std}\end{equation}
with\begin{equation}
\begin{array}{ccc}
A(z)={\displaystyle \sum_{n=1}^{\infty}\frac{z^{n}}{n\, n!}} & \textrm{and} & B(z)={\displaystyle \sum_{n=2}^{\infty}\frac{z^{n}\sum_{j=1}^{n-1}1/j}{n\, n!}}\end{array}.\label{eq:AB}\end{equation}
In Fig.%
\begin{figure}
\includegraphics[%
  width=0.80\columnwidth,
  height=0.40\columnwidth]{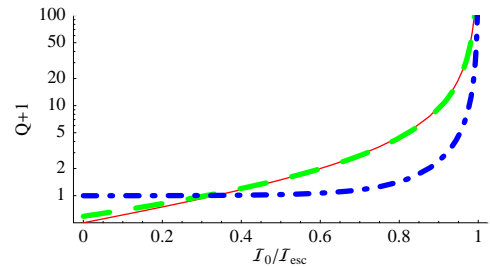}

\caption{\label{cap:Q}$Q+1$ as a function of $r=\mathcal{I}_{0}/\mathcal{I}_{esc}$.
Full line is the asymptotic line for small $\mathcal{I}_{esc}$, given
by $(1+r)/(2(1-r))$. Dashed line is for $\mathcal{I}_{esc}=1$ and
dot-dashed line is for $\mathcal{I}_{esc}=10$. Notice the logarithmic
scale. Two generic behaviors are observed: (i) if $\mathcal{I}_{0}\leq1$,
a smooth transition from sub-Poissonian to super Poissonian behavior
is observed (ii) if $\mathcal{I}_{0}>1$, a Poissonian behavior is
found until $\mathcal{I}_{0}\simeq\mathcal{I}_{esc\,}$ and then a
sharp transition from Poissonian to super-Poissonian behavior is observed.}
\end{figure}
\ref{cap:Q} we plot $Q+1$ versus $\mathcal{I}_{0}$ for three cases
where $\mathcal{I}_{esc}<1$ (escape over a shallow barrier), $\mathcal{I}_{esc}=1$
and $\mathcal{I}_{esc}>1$ (escape over a large barrier). As $\mathcal{I}_{0}\rightarrow\mathcal{I}_{esc}$,
namely the case when the chemical reaction starts close to the escaping
zone, we observe $Q\rightarrow\infty$, i.e. strong super-Poissonian
behavior. For $\mathcal{I}_{esc}\leq1$ we do not expect, and indeed
do not find an exponential behavior, since the barrier is not high.
Here, depending on initial position, either sub- or super-Poissonian
behavior is generally observed. As $\mathcal{I}_{0}$ becomes closer
to $\mathcal{I}_{esc}$, diffusion in Eq. (\ref{Eq1}) becomes more
dominant than the deterministic drift, and hence $Q$ grows. Conversely,
when $\mathcal{I}_{0}\ll\mathcal{I}_{esc}$ and $\mathcal{I}_{esc}\leq1$
drift becomes more important, leading to negative $Q$. However, diffusion
can never be neglected as it is the only mechanism leading to an eventual
escape from the well. Note that always $Q\geq-1/2$ and the absolute
minimum of $Q=-1$ is unachievable in this model. For $\mathcal{I}_{esc}>1$
and $\mathcal{I}_{0}$ sufficiently below $\mathcal{I}_{esc}$, we
have exponential behavior and $Q\simeq0$. The latter case corresponds
to Kramers' original treatment of escape from a deep metastable state. 

We now consider $\phi(\tau)$. In Laplace space and in the limit $u\gg\max\{1,1/\mathcal{I}_{0}\}$
we use \cite{A&S} to reduce Eq. (\ref{eq:Lphi-under}) to\begin{equation}
\hat{\phi}(u)\sim\left(\frac{\mathcal{I}_{esc}}{\mathcal{I}_{0}}\right)^{1/4}e^{(\mathcal{I}_{0}-\mathcal{I}_{esc})/2}e^{-2\sqrt{u}(\sqrt{\mathcal{I}_{esc}}-\sqrt{\mathcal{I}_{0}})}.\label{eq:Lpsi-large-u}\end{equation}
 The important thing to notice is the non-analytical behavior of this
asymptotic solution, namely the term of the form $e^{-\sqrt{u}}$.
This term is an indication for power law behavior (with some cutoffs)
since if $\hat{\phi}(u)=e^{-\sqrt{u}}$ for all $u$ then $\phi(\tau)\sim\tau^{-3/2}$
for large $\tau$ (e.g., \cite{Feller}). To find $\tau$ for which
$\phi(\tau)\sim\tau^{-3/2}$ we have to demand the validity of Eq.
(\ref{eq:Lpsi-large-u}) for small $\sqrt{u}(\sqrt{\mathcal{I}_{esc}}-\sqrt{\mathcal{I}_{0}})$,
while $u$ is large. This yields \begin{equation}
\left(\sqrt{\mathcal{I}_{esc}}-\sqrt{\mathcal{I}_{0}}\right)^{2}\ll\tau\ll\min\{1,\mathcal{I}_{0}\}\label{eq:IescI0<<}\end{equation}
 so that the reaction should start in the vicinity of the escaping
point. This behavior is demonstrated in Fig.%
\begin{figure}
\includegraphics[%
  width=0.80\columnwidth,
  keepaspectratio]{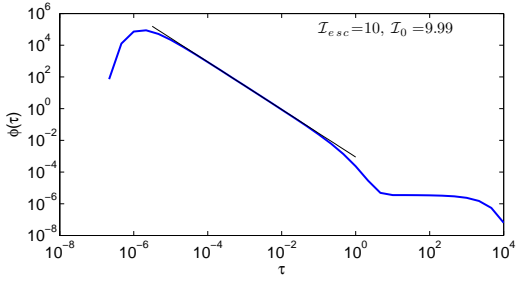}

\caption{Distribution of escape times $\phi(\tau)$ in the underdamped case.
At early and intermediate times $\phi(\tau)$ is governed by Eq. (\ref{eq:Levy-1/2}),
which exhibits a power law decay $\tau^{-3/2}$ at intermediate times
(thin line), if the initial condition $\mathcal{I}_{0}$ is close
to the escape condition $\mathcal{I}_{esc}$. At later times, there
is an exponential decay with rate given by Kramers rate.\label{cap:Underdamped-power-law}}
\end{figure}
\ref{cap:Underdamped-power-law}. 

When $\tau\rightarrow0$ we find $\phi(\tau)\rightarrow0$, a non-exponential
behavior. This behavior is due to the fact that it takes the particle
a certain amount of time to reach the boundary. More quantitatively,
for short times Eq. (\ref{eq:Lpsi-large-u}) is always a good approximation
for $\hat{\phi}(u)$, leading to \begin{eqnarray}
\phi(\tau) & \sim & \left(\frac{\mathcal{I}_{esc}}{\mathcal{I}_{0}}\right)^{1/4}e^{(\mathcal{I}_{0}-\mathcal{I}_{esc})/2}(\sqrt{\mathcal{I}_{esc}}-\sqrt{\mathcal{I}_{0}})\nonumber \\
 & \times & \frac{\exp[-(\sqrt{\mathcal{I}_{esc}}-\sqrt{\mathcal{I}_{0}})^{2}/\tau]}{\sqrt{\pi}\tau^{3/2}}.\label{eq:Levy-1/2}\end{eqnarray}

\textbf{Model 2} We turn now to the other limit of particle diffusion
- namely to the overdamped limit, and demonstrate the deviation from
Poissonian behavior. The relevant coordinate here is the spatial coordinate
\emph{x}. Restricting ourselves to one spatial dimension, the following
equation is then obtained:\begin{equation}
\frac{\partial P(x,t)}{\partial t}=(m\gamma)^{-1}\left[\frac{\partial}{\partial x}U'(x)+k_{B}T\frac{\partial^{2}}{\partial x^{2}}\right]P(x,t),\label{eq:Spatial}\end{equation}
where \emph{m} is the particle mass and $F(x)=-U'(x)$ is the deterministic
part of the force, due to potential field. We consider two cases,
when $U(x)=\pm m\omega^{2}x^{2}/2$ near the escape point, i.e., parabolic
and inverted parabolic potential, with $\omega\ll\gamma$. We define
dimensionless time $\tau=t/\theta$ and position $y=x/l$, where $\theta=\gamma\omega^{-2}$
and $l=\sqrt{k_{B}T/\left(m\omega^{2}\right)}$.

\emph{Parabolic potential.} We are interested in the FPTD from initial
position $y_{0}$ to $y_{esc}$. This function is known in Laplace
space and is given by (e.g., \cite{Siegert51})\begin{equation}
\hat{\phi}(u)=\frac{D_{-u}(-y_{0}s)e^{y_{0}^{2}/4}}{D_{-u}(-y_{esc}s)e^{y_{esc}^{2}/4}}=\frac{H_{-u}(-y_{0}s/\sqrt{2})}{H_{-u}(-y_{esc}s/\sqrt{2})},\label{eq:Lpsi-spatial}\end{equation}
where $D_{\lambda}(z)$ is the parabolic cylinder, or Weber function,
$H_{-u}(z)$ is the generalization of Hermite polynomials used in
\textsc{mathematica} and $s=\textrm{sign}(y_{esc}-y_{0})$. The PDF
$\phi(\tau)$ is normalized to 1. Using various formulas from \cite{A&S}
we can simplify Eq. (\ref{eq:Lpsi-spatial}) for large $u\gg\max\{ y_{esc}^{2},y_{0}^{2}\}$:\[
\hat{\phi}(u)\sim\exp\left[\frac{y_{0}^{2}-y_{esc}^{2}}{4}\right]e^{-\sqrt{u}|y_{esc}-y_{0}|},\]
exhibiting a non-analytical behavior of the $e^{-\sqrt{u}}$ type
similar to the previous example. In order to have the $\phi(\tau)\propto\tau^{-3/2}$
scaling we thus have to demand the validity of this approximation
for small $\sqrt{u}|y_{esc}-y_{0}|<1$. Together with the condition
$u\gg\max\{ y_{esc}^{2},y_{0}^{2}\}$ it leads to \begin{equation}
\left(y_{esc}-y_{0}\right)^{2}\ll\tau\ll\frac{1}{\max\{ y_{esc}^{2},y_{0}^{2}\}}.\label{eq:yescy0<<}\end{equation}
As we are interested in the escape over the barrier, we have either
$y_{0}<y_{esc}$ if $y_{esc}>0$, or $y_{0}>y_{esc}$ if $y_{esc}<0$.
In a deep well (low $T$), if $|y_{0}|\ll|y_{esc}|$ then $\phi(\tau)$
will be nearly exponential, similar to Model 1. If the well is not
deep, or $y_{0}$ and $y_{esc}$ are close, $Q$ shows a super Poissonian
behavior. Noticeably negative $Q$ is only possible if the initial
position of the particle is on the other side of the well from the
escape point, and high above it in energy (when $y_{0}/y_{esc}$ becomes
very negative).

\emph{Inverted parabolic potential.} The FPTD is now given by \begin{equation}
\hat{\phi}(u)={\displaystyle \frac{D_{-1-u}(-y_{0}s)}{D_{-1-u}(-y_{esc}s)}\exp\left[\frac{y_{esc}^{2}-y_{0}^{2}}{4}\right]}\label{eq:Lpsi-spatial-inverted}\end{equation}
The normalization is\[
\hat{\phi}(0)=\frac{1+\textrm{erf}(y_{0}s/\sqrt{2})}{1+\textrm{erf}(y_{esc}s/\sqrt{2})}<1.\]
The normalization is less than 1 in this case because some particles
will escape in the direction of $-s\infty$, where the potential drops
unbounded. Of course, if the particles are not allowed to escape to
infinity, so that the only exit is through $y_{esc}$ then the normalization
should be 1. However, the above formulae serve as a good approximation
for a deep well, if we consider the behavior around the parabolically
shaped escape barrier at sufficiently short times. Similar to the
previous section, for~ $u\gg\max\{1,y_{esc}^{2},y_{0}^{2}\}$\[
\hat{\phi}(u)\sim\exp\left[\frac{y_{esc}^{2}-y_{0}^{2}}{4}\right]e^{-\sqrt{u}|y_{esc}-y_{0}|}\]
and the condition for observing $\tau^{-3/2}$ scaling is\[
\left(y_{esc}-y_{0}\right)^{2}\ll\tau\ll\frac{1}{\max\{1,y_{esc}^{2},y_{0}^{2}\}}.\]

In this Letter, we demonstrated non-Poisson statistics of escape from
a potential well, which contradicts reaction rate approach used for
ensemble (bulk) dynamics. Both sub- and super-Poisson statistics were
observed, indicated by the sign of the $Q$ parameter. For small travel
distances, and for a shallow well in the overdamped case, scaling
$\phi\propto\tau^{-3/2}$ is dominant and gives rise to a strong super-Poissonian
statistics (cf. Fig. \ref{cap:Q}). The $\tau^{-3/2}$ behavior is
a result of diffusion processes \cite{Redner}. The occurrence of
such a behavior is easily understood if one realizes that in many
cases a Fokker-Planck equation with variable coefficients (e.g., Eq.
(\ref{Eq1})) can locally be approximated by an equation with constant
coefficients, i.e., by advection-diffusion equation. As travel distance
increases, characteristic time grows and drift term can compete with
diffusion, allowing for Poissonian and sub-Poissonian behavior. In
the underdamped case sub-Poissonian behavior is found in many cases
(see Fig. \ref{cap:Q}), in the overdamped case we may observe sub-Poissonian
behavior only under special conditions. As is well known, near-exponential
$\phi(\tau)$ is obtained for a deep well if initial energy is sufficiently
below the escape energy.

Finally let us compare our model results to experiments on single
molecules and to other models. Blinking nanocrystals exhibit a behavior
of occupation times of on and off times close to $\tau^{-3/2}$ with
cutoffs \cite{Brokmann03,Kuno03,Shimizu01}. The blinking is believed
to describe charging of a single nanocrystal (charged NC can be off).
Shimizu \emph{et al.} \cite{Shimizu01} briefly suggested a diffusion
in energy space to describe such behavior, however did not consider
the effect of dissipation and temperature which are always present
in case of diffusion in energy space. Recent experiments on diffusing
beads that come in and out of focus of a laser field, also exhibit
the $\tau^{-3/2}$ behavior \cite{Zumofen04}. Occupation times in
single molecule Raman experiments are described also by $\tau^{-3/2}$
behavior \cite{Bizzarri05}. Dynamics of single ion channels sometimes
exhibit the $\tau^{-3/2}$ \cite{Nadler91,Goy02}. Goychuk and H\"anggi
\cite{Goy02} suggested a model based on a reaction where space is
divided into two: a zone with free diffusion (which yields the $3/2$
law) and a zone where the reaction coordinate is climbing over a potential
field. While all these systems and models are very different, they
all exhibit a universal tendency for a $\tau^{-3/2}$ and in several
cases an exponential cutoff is observed. Thus the turn over behavior
we found, from a power law behavior, to an exponential behavior, will
be a useful concept for the single molecule domain. Further we do
not expect this behavior to be limited to Kramers problem, since diffusion
is expected to control short time dynamics of many reactions.

\begin{acknowledgments}
This work was supported by National Science Foundation grant CHE-0344930.
EB also thanks Center for Complexity Science, Israel.
\end{acknowledgments}

\begin{thebibliography}{32}
\expandafter\ifx\csname natexlab\endcsname\relax\def\natexlab#1{#1}\fi
\expandafter\ifx\csname bibnamefont\endcsname\relax
  \def\bibnamefont#1{#1}\fi
\expandafter\ifx\csname bibfnamefont\endcsname\relax
  \def\bibfnamefont#1{#1}\fi
\expandafter\ifx\csname citenamefont\endcsname\relax
  \def\citenamefont#1{#1}\fi
\expandafter\ifx\csname url\endcsname\relax
  \def\url#1{\texttt{#1}}\fi
\expandafter\ifx\csname urlprefix\endcsname\relax\def\urlprefix{URL }\fi
\providecommand{\bibinfo}[2]{#2}
\providecommand{\eprint}[2][]{\url{#2}}
 
\bibitem[{\citenamefont{Xie}(2002)}]{Xie02}
\bibinfo{author}{\bibfnamefont{X.~S.} \bibnamefont{Xie}}, \bibinfo{journal}{J.
  Chem. Phys.} \textbf{\bibinfo{volume}{117}}, \bibinfo{pages}{11024}
  (\bibinfo{year}{2002}).
 
\bibitem[{\citenamefont{Barkai et~al.}(2004)\citenamefont{Barkai, Jung, and
  Silbey}}]{BJS04}
\bibinfo{author}{\bibfnamefont{E.}~\bibnamefont{Barkai}},
  \bibinfo{author}{\bibfnamefont{Y.}~\bibnamefont{Jung}}, \bibnamefont{and}
  \bibinfo{author}{\bibnamefont{Silbey}}, \bibinfo{journal}{Annu. Rev. Phys.
  Chem.} \textbf{\bibinfo{volume}{55}}, \bibinfo{pages}{457}
  (\bibinfo{year}{2004}).
 
\bibitem[{\citenamefont{Lu et~al.}(1998)\citenamefont{Lu, Xun, and Xie}}]{Lu98}
\bibinfo{author}{\bibfnamefont{H.~P.} \bibnamefont{Lu}},
  \bibinfo{author}{\bibfnamefont{L.}~\bibnamefont{Xun}}, \bibnamefont{and}
  \bibinfo{author}{\bibfnamefont{X.~S.} \bibnamefont{Xie}},
  \bibinfo{journal}{Science} \textbf{\bibinfo{volume}{282}},
  \bibinfo{pages}{1877} (\bibinfo{year}{1998}).
 
\bibitem[{\citenamefont{Deniz et~al.}(2000)\citenamefont{Deniz, Laurence,
  Beligere, Dahan, Martin, Chemla, Dawson, Schultz, and Weiss}}]{Deniz00}
\bibinfo{author}{\bibfnamefont{A.~A.} \bibnamefont{Deniz}},
  \bibinfo{author}{\bibfnamefont{T.~A.} \bibnamefont{Laurence}},
  \bibinfo{author}{\bibfnamefont{G.~S.} \bibnamefont{Beligere}},
  \bibinfo{author}{\bibfnamefont{M.}~\bibnamefont{Dahan}},
  \bibinfo{author}{\bibfnamefont{A.~B.} \bibnamefont{Martin}},
  \bibinfo{author}{\bibfnamefont{D.~S.} \bibnamefont{Chemla}},
  \bibinfo{author}{\bibfnamefont{P.~E.} \bibnamefont{Dawson}},
  \bibinfo{author}{\bibfnamefont{P.~G.} \bibnamefont{Schultz}},
  \bibnamefont{and} \bibinfo{author}{\bibfnamefont{S.}~\bibnamefont{Weiss}},
  \bibinfo{journal}{Proc. Natl. Acad. Sci.} \textbf{\bibinfo{volume}{97}},
  \bibinfo{pages}{5179} (\bibinfo{year}{2000}).
 
\bibitem[{\citenamefont{Rhoades et~al.}(2003)\citenamefont{Rhoades,
  Gussakovsky, and Haran}}]{Rhoades03}
\bibinfo{author}{\bibfnamefont{E.}~\bibnamefont{Rhoades}},
  \bibinfo{author}{\bibfnamefont{E.}~\bibnamefont{Gussakovsky}},
  \bibnamefont{and} \bibinfo{author}{\bibfnamefont{G.}~\bibnamefont{Haran}},
  \bibinfo{journal}{Proc. Natl. Acad. Sci.} \textbf{\bibinfo{volume}{100}},
  \bibinfo{pages}{3197} (\bibinfo{year}{2003}).
 
\bibitem[{\citenamefont{Rhoades et~al.}(2004)\citenamefont{Rhoades, Cohen,
  Schuler, and Haran}}]{Rhoades04}
\bibinfo{author}{\bibfnamefont{E.}~\bibnamefont{Rhoades}},
  \bibinfo{author}{\bibfnamefont{M.}~\bibnamefont{Cohen}},
  \bibinfo{author}{\bibfnamefont{B.}~\bibnamefont{Schuler}}, \bibnamefont{and}
  \bibinfo{author}{\bibfnamefont{G.}~\bibnamefont{Haran}}, \bibinfo{journal}{J.
  Am. Chem. Soc.} \textbf{\bibinfo{volume}{126}}, \bibinfo{pages}{14686}
  (\bibinfo{year}{2004}).
 
\bibitem[{\citenamefont{Flomenbom et~al.}(2005)\citenamefont{Flomenbom,
  Velonia, Loos, Masuo, Cotlet, Engelborghs, Hofkens, Rowan, Nolte, {Van der
  Auweraer} et~al.}}]{Flomenbom05}
\bibinfo{author}{\bibfnamefont{O.}~\bibnamefont{Flomenbom}},
  \bibinfo{author}{\bibfnamefont{K.}~\bibnamefont{Velonia}},
  \bibinfo{author}{\bibfnamefont{D.}~\bibnamefont{Loos}},
  \bibinfo{author}{\bibfnamefont{S.}~\bibnamefont{Masuo}},
  \bibinfo{author}{\bibfnamefont{M.}~\bibnamefont{Cotlet}},
  \bibinfo{author}{\bibfnamefont{Y.}~\bibnamefont{Engelborghs}},
  \bibinfo{author}{\bibfnamefont{J.}~\bibnamefont{Hofkens}},
  \bibinfo{author}{\bibfnamefont{A.~E.} \bibnamefont{Rowan}},
  \bibinfo{author}{\bibfnamefont{R.~J.~M.} \bibnamefont{Nolte}},
  \bibinfo{author}{\bibfnamefont{M.}~\bibnamefont{{Van der Auweraer}}},
  \bibnamefont{et~al.}, \bibinfo{journal}{Proc. Natl. Acad. Sci.}
  \textbf{\bibinfo{volume}{102}}, \bibinfo{pages}{2368} (\bibinfo{year}{2005}).
 
\bibitem[{\citenamefont{Kramers}(1940)}]{Kramers40}
\bibinfo{author}{\bibfnamefont{H.~A.} \bibnamefont{Kramers}},
  \bibinfo{journal}{Physica} \textbf{\bibinfo{volume}{7}}, \bibinfo{pages}{284}
  (\bibinfo{year}{1940}).
 
\bibitem[{\citenamefont{H{\"a}nggi et~al.}(1990)\citenamefont{H{\"a}nggi,
  Talkner, and Borkovec}}]{Hanggi90}
\bibinfo{author}{\bibfnamefont{P.~H.} \bibnamefont{H{\"a}nggi}},
  \bibinfo{author}{\bibfnamefont{P.}~\bibnamefont{Talkner}}, \bibnamefont{and}
  \bibinfo{author}{\bibfnamefont{M.}~\bibnamefont{Borkovec}},
  \bibinfo{journal}{Reviews of modern physics} \textbf{\bibinfo{volume}{62}},
  \bibinfo{pages}{251} (\bibinfo{year}{1990}).
 
\bibitem[{\citenamefont{B{\"u}ttiker et~al.}(1983)\citenamefont{B{\"u}ttiker,
  Harris, and Landauer}}]{Buttiker83}
\bibinfo{author}{\bibfnamefont{M.}~\bibnamefont{B{\"u}ttiker}},
  \bibinfo{author}{\bibfnamefont{E.~P.} \bibnamefont{Harris}},
  \bibnamefont{and} \bibinfo{author}{\bibfnamefont{R.}~\bibnamefont{Landauer}},
  \bibinfo{journal}{Phys. Rev. B} \textbf{\bibinfo{volume}{28}},
  \bibinfo{pages}{1268} (\bibinfo{year}{1983}).
 
\bibitem[{\citenamefont{Mel'nikov and Meshkov}(1986)}]{Melnikov86}
\bibinfo{author}{\bibfnamefont{V.~I.} \bibnamefont{Mel'nikov}}
  \bibnamefont{and} \bibinfo{author}{\bibfnamefont{S.~V.}
  \bibnamefont{Meshkov}}, \bibinfo{journal}{J. Chem. Phys.}
  \textbf{\bibinfo{volume}{85}}, \bibinfo{pages}{1018} (\bibinfo{year}{1986}).
 
\bibitem[{\citenamefont{Drozdov and Brey}(1999)}]{Drozdov99}
\bibinfo{author}{\bibfnamefont{A.~N.} \bibnamefont{Drozdov}} \bibnamefont{and}
  \bibinfo{author}{\bibfnamefont{J.~J.} \bibnamefont{Brey}},
  \bibinfo{journal}{J. Chem. Phys.} \textbf{\bibinfo{volume}{110}},
  \bibinfo{pages}{2159} (\bibinfo{year}{1999}).
 
\bibitem[{\citenamefont{McCann et~al.}(1999)\citenamefont{McCann, Dykman, and
  Golding}}]{McCann99}
\bibinfo{author}{\bibfnamefont{L.~I.} \bibnamefont{McCann}},
  \bibinfo{author}{\bibfnamefont{M.}~\bibnamefont{Dykman}}, \bibnamefont{and}
  \bibinfo{author}{\bibfnamefont{B.}~\bibnamefont{Golding}},
  \bibinfo{journal}{Nature} \textbf{\bibinfo{volume}{402}},
  \bibinfo{pages}{785} (\bibinfo{year}{1999}).
 
\bibitem[{\citenamefont{Chen and Nash}(2004)}]{Chen04}
\bibinfo{author}{\bibfnamefont{L.~Y.} \bibnamefont{Chen}} \bibnamefont{and}
  \bibinfo{author}{\bibfnamefont{P.~L.} \bibnamefont{Nash}},
  \bibinfo{journal}{J. Chem. Phys.} \textbf{\bibinfo{volume}{120}},
  \bibinfo{pages}{3348} (\bibinfo{year}{2004}).
 
\bibitem[{\citenamefont{Bao and Jia}(2004)}]{Bao04}
\bibinfo{author}{\bibfnamefont{J.-D.} \bibnamefont{Bao}} \bibnamefont{and}
  \bibinfo{author}{\bibfnamefont{Y.}~\bibnamefont{Jia}},
  \bibinfo{journal}{Phys. Rev. C} \textbf{\bibinfo{volume}{69}},
  \bibinfo{pages}{027602} (\bibinfo{year}{2004}).
 
\bibitem[{\citenamefont{Shneidman}(1997)}]{Shneidman97}
\bibinfo{author}{\bibfnamefont{V.~A.} \bibnamefont{Shneidman}},
  \bibinfo{journal}{Phys. Rev. E} \textbf{\bibinfo{volume}{56}},
  \bibinfo{pages}{5257} (\bibinfo{year}{1997}).
 
\bibitem[{\citenamefont{Wang and Wolynes}(1995)}]{WangWolynes95}
\bibinfo{author}{\bibfnamefont{J.}~\bibnamefont{Wang}} \bibnamefont{and}
  \bibinfo{author}{\bibfnamefont{P.}~\bibnamefont{Wolynes}},
  \bibinfo{journal}{Phys. Rev. Lett.} \textbf{\bibinfo{volume}{74}},
  \bibinfo{pages}{4317} (\bibinfo{year}{1995}).
 
\bibitem[{\citenamefont{Chernyak et~al.}(1999)\citenamefont{Chernyak, Schulz,
  and Mukamel}}]{Chernyak99}
\bibinfo{author}{\bibfnamefont{V.}~\bibnamefont{Chernyak}},
  \bibinfo{author}{\bibfnamefont{M.}~\bibnamefont{Schulz}}, \bibnamefont{and}
  \bibinfo{author}{\bibfnamefont{S.}~\bibnamefont{Mukamel}},
  \bibinfo{journal}{J. Chem. Phys.} \textbf{\bibinfo{volume}{111}},
  \bibinfo{pages}{7416} (\bibinfo{year}{1999}).
                                                                                                                
\bibitem[{\citenamefont{Brown}(2003)}]{Brown03}
\bibinfo{author}{\bibfnamefont{F.~L.~H.} \bibnamefont{Brown}},
  \bibinfo{journal}{Phys. Rev. Lett.} \textbf{\bibinfo{volume}{90}},
  \bibinfo{pages}{028302} (\bibinfo{year}{2003}).
 
\bibitem[{\citenamefont{Weiss}(1999)}]{Weiss99}
\bibinfo{author}{\bibfnamefont{S.}~\bibnamefont{Weiss}},
  \bibinfo{journal}{Science} \textbf{\bibinfo{volume}{283}},
  \bibinfo{pages}{1676} (\bibinfo{year}{1999}).
 
\bibitem[{\citenamefont{Selvin}(2000)}]{Selvin00}
\bibinfo{author}{\bibfnamefont{P.~R.} \bibnamefont{Selvin}},
  \bibinfo{journal}{Nature Structural Biology} \textbf{\bibinfo{volume}{7}},
  \bibinfo{pages}{730} (\bibinfo{year}{2000}).
 
\bibitem[{\citenamefont{Abramowitz and Stegun}(1972)}]{A&S}
\bibinfo{author}{\bibfnamefont{M.}~\bibnamefont{Abramowitz}} \bibnamefont{and}
  \bibinfo{author}{\bibfnamefont{I.~A.} \bibnamefont{Stegun}},
  \emph{\bibinfo{title}{Handbook of mathematical functions}}
  (\bibinfo{publisher}{Dover}, \bibinfo{address}{New York},
  \bibinfo{year}{1972}).
 
\bibitem[{\citenamefont{Siegert}(1951)}]{Siegert51}
\bibinfo{author}{\bibfnamefont{A.~J.~F.} \bibnamefont{Siegert}},
  \bibinfo{journal}{Phys. Rev.} \textbf{\bibinfo{volume}{81}},
  \bibinfo{pages}{617} (\bibinfo{year}{1951}).
 
\bibitem[{\citenamefont{Feller}(1966)}]{Feller}
\bibinfo{author}{\bibfnamefont{W.}~\bibnamefont{Feller}},
  \emph{\bibinfo{title}{An introduction to probability theory and its
  applications}}, vol.~\bibinfo{volume}{2} (\bibinfo{publisher}{Wiley\&Sons},
  \bibinfo{address}{New York}, \bibinfo{year}{1966}).
 
\bibitem[{\citenamefont{Redner}(2001)}]{Redner}
\bibinfo{author}{\bibfnamefont{S.}~\bibnamefont{Redner}},
  \emph{\bibinfo{title}{A Guide to First-Passage Processes}}
  (\bibinfo{publisher}{Cambridge University Press}, \bibinfo{address}{New
  York}, \bibinfo{year}{2001}).
 
\bibitem[{\citenamefont{Brokmann et~al.}(2003)\citenamefont{Brokmann, Hermier,
  Messin, Desbiolles, Bouchaud, and Dahan}}]{Brokmann03}
\bibinfo{author}{\bibfnamefont{X.}~\bibnamefont{Brokmann}},
  \bibinfo{author}{\bibfnamefont{J.~P.} \bibnamefont{Hermier}},
  \bibinfo{author}{\bibfnamefont{G.}~\bibnamefont{Messin}},
  \bibinfo{author}{\bibfnamefont{P.}~\bibnamefont{Desbiolles}},
  \bibinfo{author}{\bibfnamefont{J.-P.} \bibnamefont{Bouchaud}},
  \bibnamefont{and} \bibinfo{author}{\bibfnamefont{M.}~\bibnamefont{Dahan}},
  \bibinfo{journal}{Phys. Rev. Lett.} \textbf{\bibinfo{volume}{90}},
  \bibinfo{pages}{120601} (\bibinfo{year}{2003}).
 
\bibitem[{\citenamefont{Kuno et~al.}(2003)\citenamefont{Kuno, Fromm, Johnson,
  Gallagher, and Nesbitt}}]{Kuno03}
\bibinfo{author}{\bibfnamefont{M.}~\bibnamefont{Kuno}},
  \bibinfo{author}{\bibfnamefont{D.~P.} \bibnamefont{Fromm}},
  \bibinfo{author}{\bibfnamefont{S.~T.} \bibnamefont{Johnson}},
  \bibinfo{author}{\bibfnamefont{A.}~\bibnamefont{Gallagher}},
  \bibnamefont{and} \bibinfo{author}{\bibfnamefont{D.~J.}
  \bibnamefont{Nesbitt}}, \bibinfo{journal}{Phys. Rev. B}
  \textbf{\bibinfo{volume}{67}}, \bibinfo{pages}{125304}
  (\bibinfo{year}{2003}).
 
\bibitem[{\citenamefont{Shimizu et~al.}(2001)\citenamefont{Shimizu, Neuhauser,
  Leatherdale, Empedocles, Woo, and Bawendi}}]{Shimizu01}
\bibinfo{author}{\bibfnamefont{K.~T.} \bibnamefont{Shimizu}},
  \bibinfo{author}{\bibfnamefont{R.~G.} \bibnamefont{Neuhauser}},
  \bibinfo{author}{\bibfnamefont{C.~A.} \bibnamefont{Leatherdale}},
  \bibinfo{author}{\bibfnamefont{S.~A.} \bibnamefont{Empedocles}},
  \bibinfo{author}{\bibfnamefont{W.~K.} \bibnamefont{Woo}}, \bibnamefont{and}
  \bibinfo{author}{\bibfnamefont{M.~G.} \bibnamefont{Bawendi}},
  \bibinfo{journal}{Phys. Rev. B} \textbf{\bibinfo{volume}{63}},
  \bibinfo{pages}{205316} (\bibinfo{year}{2001}).
 
\bibitem[{\citenamefont{Zumofen et~al.}(2004)\citenamefont{Zumofen, Hohlbein,
  and H{\"u}bner}}]{Zumofen04}
\bibinfo{author}{\bibfnamefont{G.}~\bibnamefont{Zumofen}},
  \bibinfo{author}{\bibfnamefont{J.}~\bibnamefont{Hohlbein}}, \bibnamefont{and}
  \bibinfo{author}{\bibfnamefont{C.~G.} \bibnamefont{H{\"u}bner}},
  \bibinfo{journal}{Phys. Rev. Lett.} \textbf{\bibinfo{volume}{93}},
  \bibinfo{pages}{260601} (\bibinfo{year}{2004}).
 
\bibitem[{\citenamefont{Bizzarri and Cannistraro}(2005)}]{Bizzarri05}
\bibinfo{author}{\bibfnamefont{A.~R.} \bibnamefont{Bizzarri}} \bibnamefont{and}
  \bibinfo{author}{\bibfnamefont{S.}~\bibnamefont{Cannistraro}},
  \bibinfo{journal}{Phys. Rev. Lett.} \textbf{\bibinfo{volume}{94}},
  \bibinfo{pages}{068303} (\bibinfo{year}{2005}).
 
\bibitem[{\citenamefont{Nadler and Stein}(1991)}]{Nadler91}
\bibinfo{author}{\bibfnamefont{W.}~\bibnamefont{Nadler}} \bibnamefont{and}
  \bibinfo{author}{\bibfnamefont{D.~L.} \bibnamefont{Stein}},
  \bibinfo{journal}{Proc. Natl. Acad. Sci.} \textbf{\bibinfo{volume}{88}},
  \bibinfo{pages}{6750} (\bibinfo{year}{1991}).
 
\bibitem[{\citenamefont{Goychuk and H{\"a}nggi}(2002)}]{Goy02}
\bibinfo{author}{\bibfnamefont{I.}~\bibnamefont{Goychuk}} \bibnamefont{and}
  \bibinfo{author}{\bibfnamefont{P.}~\bibnamefont{H{\"a}nggi}},
  \bibinfo{journal}{Proc. Natl. Acad. Sci.} \textbf{\bibinfo{volume}{99}},
  \bibinfo{pages}{3552} (\bibinfo{year}{2002}).
 
\end{thebibliography}

\end{document}